\documentclass[11pt, a4paper]{article}
\pdfoutput = 1
\usepackage[height=22cm,width=16cm, centering]{geometry}

\usepackage{setspace}

\usepackage[utf8]{inputenc}
\usepackage[sort&compress, numbers, merge]{natbib}

\usepackage{amsmath, amssymb, mathrsfs,  comment}
\usepackage{ifpdf, xcolor}

\ifpdf        
  \usepackage{graphicx}     
  \usepackage[bookmarksopen,colorlinks=true, linkcolor=link_col,
  citecolor=cite_col, urlcolor=url_col,linktocpage=false]{hyperref}
\else     
  \usepackage[dvipdfmx]{graphicx}     
  \usepackage[dvipdfmx,bookmarksopen,colorlinks=true,linkcolor=link_col,
  citecolor=cite_col ,urlcolor=url_col,linktocpage=false]{hyperref}
\fi

\usepackage{multicol}
\definecolor{link_col}{rgb}{0.0, 0, 0.8}
\definecolor{cite_col}{rgb}{0.6, 0, 0.3}
\definecolor{url_col}{rgb}{0.6, 0, 0.3}

\def\Mpl{M_{\rm P}}

\begin{document}

\begin{titlepage}
\begin{center}
\leavevmode \\

\vskip -2 mm
{\small 
\hfill IPMU17-0116\\
\hfill UT-17-28\\
\hfill KEK-TH-1992\\
\hfill KEK-Cosmo-210\\
}

\noindent
\vskip 1.5 cm
{\huge Gravitino Problem in Inflation Driven by \\ \vskip 1mm Inflaton-Polonyi K\"ahler Coupling}

\vskip 0.8 cm

{\Large Fuminori Hasegawa$^{1,2}$,  
 Kazunori Nakayama$^3$,\\ Takahiro Terada$^4$, and Yusuke Yamada$^5$}

\vskip 1. cm

{\textit {\small
$^1$Institute for Cosmic Ray Research, The University of Tokyo, Kashiwa, Chiba 277-8582, Japan\\
$^2$Kavli IPMU (WPI), UTIAS, The University of Tokyo, Kashiwa, Chiba 277-8583, Japan\\
$^3$Department of Physics, Faculty of Science, The University of Tokyo, \\ Bunkyo-ku, Tokyo 113-0033, Japan\\
$^4$Theory Center, IPNS, KEK, Tsukuba, Ibaraki 305-0801, Japan\\
$^5$Stanford Institute for Theoretical Physics and Department of Physics,\\
Stanford University, Stanford, CA 94305, U.S.A.}}

\vskip 1.6 cm

\begin{abstract}
{\normalsize 
We discuss the cosmological gravitino problem in inflation models in which the inflaton potential is constructed from K\"{a}hler potential rather than superpotential: a representative model is $\overline{\text{D}3}$-induced geometric inflation.
A critical ingredient in this type of models is the coupling of the inflaton and Polonyi (supersymmetry-breaking) field in the K\"{a}hler potential, which is needed to build the inflaton potential. 
We point out the same coupling let the inflaton dominantly decay into a pair of inflatino and gravitino causing the gravitino problem. We propose some possible solutions to this problem.}
\end{abstract}

\end{center}

\end{titlepage}

\section{Introduction}

It is important to examine how an inflation model is embedded into a motivated theory beyond the Standard Model such as supersymmetry (SUSY) or supergravity, and to study consequences of the embedding.
For this purpose, it is useful to study production of long-lived particles in the reheating epoch after inflation because overproduction of such particles is constrained by cosmological observations.  Dark matter abundance~\cite{Ade:2015xua} and Big-Bang Nucleosynthesis (BBN)~\cite{Moroi:1995fs, Jedamzik:2004er, Kawasaki:2004yh, Kawasaki:2004qu, Jedamzik:2006xz, Kawasaki:2008qe} provide stringent constraints.

For a given inflation model, there are several ways to embed it into supergravity.
One can reconstruct the same inflation potential by several implementation methods.
For example, one can realize an arbitrary positive semidefinite potential with a so-called stabilizer field~\cite{Kallosh:2010ug, Kallosh:2010xz, Kawasaki:2000yn}, without the stabilizer field~\cite{Ketov:2014hya, Ketov:2014qha, Ketov:2016gej, Ferrara:2016vzg, Roest:2015qya, Linde:2015uga}, with a nilpotent field~\cite{DallAgata:2014qsj, Carrasco:2015pla, Ferrara:2014kva}, with $D$-term~\cite{Ferrara:2013rsa, Farakos:2013cqa, Binetruy:1996xj, Halyo:1996pp}, and with ``geometric K\"{a}hler corrections''~\cite{McDonough:2016der, Kallosh:2017wnt} discussed below.

In this short note, we point out that there is a significant constraint on the class of models of $F$-term inflation in which the inflaton potential is constructed from non-minimal coupling between the inflaton $\phi$ and the SUSY breaking field $X$ in the K\"{a}hler potential\footnote{
In reality, other matter fields enter $K$ and $W$.  We assume these fields do not play a major role in SUSY breaking.  It is usually the case that SUSY Standard Model fields communicate with $X$ indirectly, and their coupling to the inflaton should not be too large [see discussion below eq.~\eqref{coupling_spike}].  In this way, our analysis based on the two superfields $\phi$ and $X$ is not affected significantly by addition of the matter fields.
}:
\begin{align*}
	&K = k(\phi,\bar\phi) + h(\phi,\bar\phi) |X|^2, \\
	&W = \mu^2 X + W_0.
\end{align*}
That is, we will see that a large amount of gravitinos are produced (gravitino problem~\cite{Khlopov:1984pf,Ellis:1984eq}) by the inflaton decay, and this sets a constraint on the relation between the inflaton mass and the gravitino mass. 
In this class of models, the scalar potential is governed solely by K\"ahler potential $K$ in contrast to the conventional model building, where both $K$ and superpotential $W$ are responsible.
Also, $X$ is responsible for SUSY breaking from the inflationary epoch to the present epoch.
The new model building emerged in generalizing inflation in supergravity utilizing a nilpotent superfield~\cite{Antoniadis:2014oya, Ferrara:2014kva, Kallosh:2014via, Aoki:2014pna, Dall'Agata:2014oka, Kallosh:2014hxa, Scalisi:2015qga, Ferrara:2015tyn, Dall'Agata:2015lek, Kallosh:2016hcm, Dall'Agata:2016yof, McDonough:2016der, Argurio:2017joe}, which is a nonlinear realization of SUSY~\cite{Rocek:1978nb, Komargodski:2009rz}. It has also been pointed out that the nilpotent superfield is an effective description of anti-D3 brane ($\overline{\text{D}3}$) in superstring theory~\cite{Ferrara:2014kva,Kallosh:2014wsa,Kallosh:2015nia}. From such a viewpoint, one can regard the non-minimal coupling $h(\phi,\bar\phi)|X|^2$ as the back-reaction to the brane from Calabi-Yau moduli, that is, the inflation potential has a beautiful geometric origin. One can realize various inflaton potentials \textit{e.g.}~the $\alpha$-attractor~\cite{Kallosh:2013yoa, Galante:2014ifa, Carrasco:2015pla} by choosing $k$ and $h$~\cite{McDonough:2016der, Kallosh:2017wnt}. The similar setup can be realized even if $X$ is not nilpotent although the meaning of such a model in string theory is unclear.
In the following discussion, however, whether it is related to $\overline{\text{D}3}$ or whether the SUSY breaking field is nilpotent is not essential.

In this letter we study the nonthermal gravitino production in this class of models.
A series of works studied the nonthermal gravitino production from both the perturbative inflaton decay~\cite{Endo:2006zj,Nakamura:2006uc,Asaka:2006bv,Dine:2006ii,Endo:2006tf,Kawasaki:2006hm,Endo:2007ih,Endo:2007sz,Ema:2016oxl}  and the non-perturbative one~\cite{Kallosh:1999jj, Giudice:1999yt, Giudice:1999am, Kallosh:2000ve, Nilles:2001ry, Nilles:2001fg, Ema:2016oxl}. 
A crucial difference of the present model from those studied previously is that the inflaton mass comes from the SUSY breaking effect
and typically the inflatino, the fermionic superpartner of the inflaton, is much lighter than the inflaton.
Hence the inflaton efficiently decays into the gravitino plus inflatino, leading to the severe gravitino problem.

We use the reduced-Planck unit $M_{\text{P}}=1/\sqrt{8\pi G} = 1$ unless explicitly written.

\section{Inflation from K\"{a}hler corrections} 

\subsection{Model}

Let us consider the following model~\cite{McDonough:2016der, Kallosh:2017wnt},
\begin{align}
	&K = k(\phi,\bar{\phi}) + h(\phi,\bar{\phi}) |X|^2,  \label{K} \\
	&W = \mu^2 X + W_0  \label{W},
\end{align}
where $\phi$ is the inflaton superfield, and $X$ is the superfield responsible for SUSY breaking.  The parameters $\mu^2$ and $W_0$ are taken to be real without loss of generality. In Ref.~\cite{Kallosh:2017wnt},  $X$-dependence is solely in $K$, \textit{i.e.}~$K= k(\phi, \bar{\phi}) + X + \bar{X} + h(\phi, \bar{\phi}) |X|^2$ and $W=W_0$ (const.), but we have transformed it to the above form by K\"{a}hler-Weyl transformation and redefinition of the normalization of $X$ and $h$.

Also, $X$ is taken to be nilpotent ($X^2=0$) in Refs.~\cite{McDonough:2016der, Kallosh:2017wnt}, but whether it is nilpotent or not is irrelevant in our discussion.  Even if $X$ is not nilpotent, an additional term like $\Delta K = - |X|^4/\Lambda^2$ with $\Lambda$ some mass scale strongly stabilizes $X$ around its origin, and then, the similar model building is possible.\footnote{
See Refs.~\cite{Ghilencea:2015aph, Dudas:2016eej, Argurio:2017joe} for restrictions on the possible range of $\Lambda$ and the viability of the nilpotency condition.
}  For definiteness, we take $X$ nilpotent in the following presentation.

The scalar potential is
\begin{align}
	V(\phi,\bar{\phi})= e^k \left[ \frac{\mu^4}{h} + \left( \frac{|k_\phi|^2}{k_{\phi\bar\phi}}-3 \right)W_0^2 \right],  \label{V}
\end{align}
where we have set $X=0$ due to the nilpotency, and we adopt the notation like $k_\phi = \partial k/\partial \phi$ and $k_{\phi\bar{\phi}}= \partial^2 k/\partial \phi \partial \bar{\phi}$ etc.
We assume that $K$ is shift-symmetric with respect to the inflaton and that the scalar partner of the inflaton (sinflaton) settles down to its minimum during inflation.  We also assume that $k_\phi=0$ is realized dynamically during and after inflation.\footnote{$k_{\phi}=0$ is easily realized if the sinflaton has a $Z_2$ symmetry. 
Meanwhile, nonzero $k_{\phi}$ is another important source for the gravitino production.}  The constant value of $k$ during inflation can be absorbed by the redefinition of the normalization of the superpotential, so we can set $e^k = 1$ throughout the inflation trajectory.  
As proposed in Ref.~\cite{Kallosh:2017wnt}, if we take
\begin{align}
h(\phi, \bar{\phi}) = & \frac{3 m_{3/2}^2}{{\bf V}(\phi,\bar{\phi}) + 3 m_{3/2}^2},
\end{align}
the scalar potential becomes
\begin{align}
	V(\phi,\bar{\phi})=  {\bf V}(\phi, \bar{\phi}), 
\end{align}
where we have used the condition of the vanishing cosmological constant $\mu^2 = \sqrt{3} m_{3/2}$.  
Thus, one can implement an arbitrary inflaton potential.

\subsection{Mass spectrum}

The inflaton mass $m_{\phi}$ is a free parameter in this type of models because the inflaton potential is arbitrary.
For typical chaotic/large-field inflation models, the inflaton mass is of order $10^{13}$ GeV from the normalization of the CMB anisotropy.
On the other hand, the inflatino mass is not necessarily related to $m_\phi$ because the inflaton potential does not originate from the superpotential. This is a characteristic feature of this type of inflation models. The mass of the inflatino tends to be of order the gravitino mass $m_{3/2}$.

We now see this concretely.
SUSY is spontaneously broken by the $F$-term of $X$ and the Goldstino $\widetilde X$ is absorbed by the gravitino (super-Higgs mechanism).\footnote{
In the oscillating inflaton background, the kinetic energy of the inflaton also contributes to SUSY breaking~\cite{Kallosh:2000ve}.  This effect is negligible when $H\ll m_{3/2}$, where $H$ denotes the Hubble parameter. 
}
  The gravitino mass is given by $m_{3/2}=e^{K/2}W=W_0$.
The inflatino mass parameter $m_{\widetilde{\phi}}$ at the minimum of the potential is given by 
\begin{align}
m_{\widetilde{\phi}}=& m_{3/2} \left( k_{\phi\phi} - (k_{\phi\bar{\phi}})^{-1}k_{\phi\phi\bar{\phi}}k_{\phi} + \frac{1}{3} (k_{\phi})^2 \right) .
\end{align}
For example, let us take $k= - \frac{1}{2} (\phi - \bar\phi)^2$. It leads to $m_{\widetilde{\phi}}= - m_{3/2}$. The physical inflatino mass is $|m_{\widetilde{\phi}}|=m_{3/2}$.
On the other hand, for the minimal K\"ahler potential $k=|\phi|^2$, the inflatino is massless and acts as dark radiation 
if it is produced in the early universe.
Hereafter we assume that the inflatino has a mass of the order of the gravitino,
which is typical in the class of models in which $k=k_\phi=0$ along the inflaton trajectory.

The mass of the sinflaton is more model-dependent because one can give it a mass by hand in ${\bf V}(\phi, \bar{\phi})$.  
If ${\bf V}(\phi, \bar{\phi})$ is independent of sinflaton, the sinflaton mass as well as the inflatino mass tends to be of the order of the gravitino mass. 
If its oscillation amplitude is induced by the scalar dynamics after inflation, it can potentially become a source of the moduli problem~\cite{Coughlan:1983ci, Banks:1993en, deCarlos:1993wie} depending on its couplings (decay channels) to matter fields.  For more generic choices of ${\bf V}$, the sinflaton can be heavier, and the moduli problem can be avoided.

\section{Single gravitino production}  

Now let us study the gravitino production from the inflaton decay in this model.
We make use of the Goldstino picture hereafter. 
Since the SUSY breaking in the present universe is carried by the $F$-term of $X$, we identify $\widetilde X$ as the Goldstino.
Also note that the (transverse) gravitino mass $m_{3/2}$ is constant throughout the whole history of the universe.
From the non-minimal K\"ahler term (\ref{K}), we have the following inflaton-inflatino-Goldstino interaction:
\begin{align}
	\mathcal L = \frac{\sqrt 3 m_{3/2} \Mpl h_\phi}{h} \widetilde\phi \widetilde X + {\rm h.c.}
	\simeq \frac{\sqrt 3 m_{3/2} \Mpl h_{\phi\bar\phi}}{k_{\phi\bar\phi} h^{3/2}} \phi_{\text{c}}^*\widetilde\phi_{\text{c}} \widetilde X_{\text{c}} + {\rm h.c.}, \label{Lint}
\end{align}
where the subscript c represents a canonically normalized field.\footnote{
	We approximated $\phi_{\text{c}} \simeq \sqrt{k_{\phi\bar\phi}} \phi$, which may be justified around the potential minimum $\phi=0$. Note that this field $\phi$ should be regarded as the fluctuation around the vacuum expectation value.
}
In the last similarity we assumed\footnote{If $|h_{\phi\bar\phi}| \lesssim  |h_{\phi\phi}|$, there appears a tachyonic mode unless additional couplings \textit{e.g.}~to matters are introduced.  Even if $|h_{\phi\bar\phi}| \sim  |h_{\phi\phi}|$, the following argument is modified only by an order one factor.}
 $|h_{\phi\bar\phi}| \gg  |h_{\phi\phi}|$ and neglected $|h_\phi|^2$ because it vanishes at the minimum. 
On the other hand, the same K\"ahler potential yields the (SUSY breaking) inflaton mass as
\begin{align}
	m_{\phi}^2 \simeq -\frac{3m _{3/2}^2 \Mpl^2 h_{\phi\bar\phi}}{k_{\phi\bar\phi}h^2} \simeq \frac{\textbf{V}_{\phi\bar{\phi}}}{k_{\phi\bar{\phi}}}.
\end{align}
Note that if ${\bf V}(\phi, \bar{\phi})$ does not contain quadratic term but terms with higher power in $\phi$,
$m_{\phi}^2$ is $\phi$-dependent and should be regarded as the ``effective'' mass.
Then eq.~(\ref{Lint}) can be written as
\begin{align}
	\mathcal L \simeq -\frac{m_{\phi}^2 h^{1/2}}{\sqrt 3 m_{3/2} \Mpl} \phi_{\text{c}}^*\widetilde\phi_{\text{c}} \widetilde X_{\text{c}} + {\rm h.c.}
	\simeq - \frac{m_\phi^2}{\sqrt {3 m_{3/2}^2 \Mpl^2 + {\bf V} }} \phi_{\text{c}}^*\widetilde\phi_{\text{c}} \widetilde X_{\text{c}} + {\rm h.c.}  \label{Ldec}
\end{align}
The square root of the above Yukawa coupling is the ratio between the inflaton mass and the unitarity bound scale for the gravitino scattering which is expected in the cosmological background~\cite{Casalbuoni:1988sx, Kallosh:2000ve, Dall'Agata:2014oka, Kahn:2015mla, Ferrara:2015tyn, Carrasco:2015iij, Delacretaz:2016nhw}. To avoid the strong coupling, we require that the ratio is smaller than $4\pi$.

As a first example, let us take ${\bf V}(\phi, \bar{\phi})\simeq k_{\phi\bar\phi} m_\phi^2|\phi|^2$, \textit{i.e.}~$m_{\phi}$ is constant.\footnote{
	The inflaton potential should be modified from the quadratic one to fit the Planck observation~\cite{Ade:2015lrj},
	but it is irrelevant in the present discussion because we are interested in the potential around the minimum $\phi=0$.
} 
In this case, the condition that we are not in the strong coupling regime leads to
\begin{align}
m_{3/2} \gtrsim 2 \times 10^6 \, \text{GeV} \left( \frac{m_{\phi}}{10^{13}\,\text{GeV}} \right)^2. \label{m_unitarity}
\end{align}
This already implies that the large hierarchy between the masses of the inflaton and the gravitino has difficulty. 
In fact, it is noticed in Ref.~\cite{McDonough:2016der} that the weak coupling regime leads to high scale SUSY breaking to fit the Planck data.
In the late-time limit $H\ll m_{3/2}$, 
we can safely take $h\simeq 1$ in eq.~(\ref{Ldec}) and the well-known result for the decay width of the
heavy SUSY particle into the light gravitino is recovered~\cite{Moroi:1995fs, Buchmuller:2004rq}:
\begin{align}
	\Gamma (\phi \to \widetilde{\phi} \psi_{3/2}) \simeq \frac{m_{\phi}^5}{48 \pi m_{3/2}^2 M_{\text{P}}^2},   \label{decayrate}
\end{align}
where we used $m_{\phi} \gg |m_{\widetilde{\phi}}|, m_{3/2}$.
The rate is enhanced by $(m_{\phi}/m_{3/2})^2$ compared to the Planck-suppressed rate $\sim m_{\phi}^3/M_{\text{P}}^2$. We emphasize this is only true for the case of a large mass difference between a boson $\phi$ and a fermion $\widetilde \phi$, and it is originated in our model by the non-minimal coupling $h(\phi,\phi^{\dag})|X|^2$ in the K\"{a}hler potential, which is the essential ingredient of the K\"{a}hler-induced models of inflation.

In the opposite regime $H \gg m_{3/2}$, the situation is more complicated, since the interaction (\ref{Ldec}) is written as
\begin{align}
	\mathcal L =- \frac{m_\phi^2 \phi_{\text{c}}^*}{\sqrt {3 m_{3/2}^2 \Mpl^2 + m_\phi^2|\phi_{\text{c}}|^2 }}\widetilde\phi_{\text{c}} \widetilde X_{\text{c}} + {\rm h.c.}, \label{coupling_spike}
\end{align}
and this overall coefficient shows a spike-like behavior as a function of time during the inflaton oscillation.
The particle production with such a spike-like mass term was studied in Refs.~\cite{Amin:2015ftc,Ema:2016dny} 
and there efficient particle production is found to exist.\footnote{
	The momentum of the produced gravitino/inflatino due to the spike is about  $k\sim m_\phi H/m_{3/2}$ and it can be much higher than the unitarity bound scale
	even if the constraint (\ref{m_unitarity}) is satisfied.
}
We do not go into details of this analysis in this letter, but for our purpose it is enough to notice that the perturbative decay rate (\ref{decayrate}), which is applicable for $H\lesssim m_{3/2}$, 
is already so huge, \textit{i.e.} $\Gamma \gg m_{3/2} \gtrsim H$, that the inflaton completely decays into the gravitino and the inflatino unless the inflaton is very light or the gravitino is superheavy.
Therefore, the universe is likely dominated by the inflatino and gravitino at $H\sim m_{3/2}$ at latest.  
It might be possible to introduce large couplings of the inflaton to the SUSY Standard Model sector to reduce the branching ratio to the gravitino-inflatino pair, but then the reheating temperature will be so high that thermal production of the gravitino becomes problematic.

Since the universe is dominated by the gravitino and inflatino, 
Standard Model particles must be reheated through their decay.
The decay rate of the inflatino depends on unspecified inflaton-Standard Model matter couplings and it can decay earlier than the gravitino.
As far as the inflatino mass is comparable to the gravitino, the universe is likely to be finally dominated by the gravitino.
The reheating temperature after the gravitino decay is estimated as
\begin{align}
	T_{\rm R} \simeq 0.1\,{\rm GeV} \left( \frac{m_{3/2}}{10^6\,{\rm GeV}} \right)^{3/2}. 
\end{align}
In our setup, the gravitino and inflatino are so heavy [see eq.~\eqref{m_unitarity}] that they themselves are harmless for BBN.
Although the reheating temperature can be higher than the freeze-out temperature of the lightest SUSY particles (LSP),
the LSP is typically too heavy to be consistent with observed dark matter abundance.
The LSP overproduction problem can be avoided in several ways by introducing small $R$-parity violation, 
or by assuming some late-time entropy production processes.
It may also be possible to make the LSP extraordinary light with some elaborated mechanism of the mediation of SUSY breaking.
See Ref.~\cite{Jeong:2012en} for more details of the gravitino-dominated universe. 
In any case, both the unitarity/strong-coupling issue and the gravitino problem imply that SUSY is broken at a high energy scale.

We also make a comment on the case of a more general inflaton potential.
It may be possible to suppress the quadratic term in ${\bf V}(\phi, \bar{\phi})$ by hand so that the inflaton is (nearly) massless around the potential minimum
and its decay into the gravitino-inflatino pair is kinematically forbidden.
It should be noticed that still there is a severe gravitino problem.
This is because the inflaton necessarily obtains an effective SUSY breaking mass whatever the potential is,
which again induces the gravitino production.
As an example, let us suppose ${\bf V}(\phi, \bar{\phi}) \simeq \lambda|\phi_{\text{c}}|^4$ around the minimum.
Then the ``effective'' inflaton mass during the oscillation is $m_\phi^2 \sim \lambda \Phi^2$  with $\Phi$ being the oscillation amplitude.
This effective mass is also the SUSY breaking mass, so the gravtino production rate is again given by eq.~(\ref{decayrate}) up to $\mathcal O(1)$ numerical factor,
once $m_\phi$ is interpreted as this effective mass.
A difference is that $m_\phi$ is decreasing with time due to the Hubble expansion
and the production is accessible only for $H \gtrsim m_{3/2}^2/(\sqrt\lambda \Mpl)$.
For $H\gg m_{3/2}$, eq.~(\ref{Ldec}) also shows a spike-like behavior and the efficient gravitino-inflatino production is expected.
In this way, the case with a suppressed vacuum inflaton mass will also lead to the universe dominated by the gravitino and inflatino.

\section{Discussion and Summary} \label{sec:dis}

We have seen that the type of inflation models recently advocated in the context of inflaton-nilpotent field coupling has the cosmological gravitino problem. The universe is dominated by the gravitino and the inflatino.   This is because the coupling between the inflaton and the SUSY breaking sector induces the enhanced soft SUSY breaking term, and hence the hierarchical mass difference between the boson and the fermion.
We used a nilpotent field $X$ for definiteness, but the same is true for an unconstrained chiral superfield which breaks SUSY.

As we saw above, the large gravitino mass and the reheating by gravitino decay may be a viable solution, but let us consider ways to avoid the enhanced gravitino production itself.
The core of the problem is the separation of the inflaton mass and the gravitino mass ($\sim$ the inflatino mass).
Therefore, degeneracy of the two scales solves the problem.
We discussed that simply lowering the inflaton mass does not solve the problem because of the effective mass.
Still, one can raise the SUSY breaking scale close to the (effective) inflaton mass.
One has to be aware of the fact that even a small amount of gravitino leads to too many LSPs, $n_{3/2}/s \lesssim 5 \times 10^{-15} \times (10^5 \, \text{GeV}/m_{\text{LSP}})$, where $n_{3/2}$ and $s$ are the gravitino number density and entropy density, respectively. 
One may also consider a scenario in which the gravitino mass is larger than the inflaton mass and the reheating temperature~\cite{Benakli:2017whb, Dudas:2017rpa}.  Then, the inflaton cannot decay into gravitinos.
In this way, high scale SUSY breaking seems to be the typical solution to the gravitino problem in the considered models.  This automatically solves the moduli problem mentioned above too.

The high scale SUSY breaking solution may be reasonable from the viewpoint of string theory. Since the $\overline{\text{D}3}$ geometric inflation would be embedded into string theory, we need to consider the stabilization of extra spaces as well. In the known moduli stabilization models such as KKLT~\cite{Kachru:2003aw} and LVS~\cite{Balasubramanian:2005zx}, the gravitino mass scale tends to correlate to the Calabi-Yau volume modulus. In such cases, to avoid the decompactification/overshooting problem, the inflation scale is required to be smaller than the gravitino mass scale unless we require fine-tunings of parameters~\cite{Kallosh:2004yh,Conlon:2008cj}. Hence, the high scale SUSY solution to the gravitino problem seems compatible with string theory realization of the $\overline{\text{D}3}$ geometric inflation.

One of the solutions is to consider (more conventional) models with a SUSY mass for the inflaton supermultiplet, \textit{e.g.}
\begin{align}
W (\phi, X) = & f(\phi) X + g(\phi),
\end{align}
with or without the non-minimal coupling $h(\phi, \bar{\phi})\neq 1$ in the K\"{a}hler potential.
The above superpotential is the general one for the nilpotent $X$, but one can consider more general $X$-dependence for the unconstrained $X$ as long as it generates a suitable inflation potential.  In this setup, the decay channel considered here can be kinematically forbidden or suppressed~\cite{Terada:2014uia}. If the inflatino degenerates with the inflaton ($m_{\phi} \simeq m_{\widetilde{\phi}}$) due to the SUSY mass, the decay rate is estimated as
\begin{align}
\Gamma (\phi \to \widetilde{\phi}\psi_{3/2}) \simeq \frac{m_{\phi}^{3}}{3\pi M_{\text{P}}^{2}}\left( \frac{m_{3/2}}{m_{\phi}} \right)^{2}\Delta (\Delta^{2}-1)^{\frac{3}{2}},
\end{align}
where $\Delta \equiv (m_\phi - |m_{\widetilde{\phi}}|)/m_{3/2}$ parametrizes the mass difference.

In summary, if we stick to the K\"{a}hler potential-based models of inflation [eqs.~\eqref{K} and \eqref{W}], there should not be a large mass hierarchy between the inflaton and the gravitino. 
This is because otherwise the universe is dominated by the gravitino and inflatino from the inflaton decay, which is a cosmological disaster unless some extra assumptions are made.

\section*{Acknowledgments}
We are grateful to Evan McDonough, Kyohei Mukaida and Marco Scalisi for discussion. YY would like to thank Renata Kallosh, Andrei Linde and Diederik Roest for collaboration in Ref.~\cite{Kallosh:2017wnt} and valuable discussion and comments.
This work is supported in part by JSPS Research Fellowship for Young Scientists  [FH, TT], JSPS Grant-in-Aid for Scientific Research on Scientific Research A (JP26247042), Young Scientists B (JP26800121) and Innovative Areas (JP26104009, JP15H05888 and JP17H06359) [KN], JSPS KAKENHI grant Number JP17J00731 [TT], SITP and the NSF grant PHY-1720397 [YY].

\small

\bibliographystyle{utphys}
\bibliography{single_gravitino}

\end{document}